%% file: main.tex
\documentclass{svproc}

\usepackage{url}
\usepackage{amsmath,amssymb,amsfonts}
\usepackage{algorithmic}
\usepackage{algorithm}
\usepackage{graphicx}
\usepackage{textcomp}
\usepackage{xcolor}
\usepackage{array}
\usepackage{multirow}
\usepackage{enumitem}
\usepackage{tikz}
\usetikzlibrary{shapes.geometric, arrows, positioning}
\usepackage{pifont}
\usepackage{placeins}
\usepackage[hidelinks]{hyperref}

\setlist[itemize]{label=\textbullet, leftmargin=*}

\newcommand{\PreserveBackslash}[1]{\let\temp=\\#1\let\\=\temp}
\newcolumntype{C}[1]{>{\PreserveBackslash\centering}p{#1}}
\newcolumntype{R}[1]{>{\PreserveBackslash\raggedleft}p{#1}}
\newcolumntype{L}[1]{>{\PreserveBackslash\raggedright}p{#1}}

\begin{document}
\mainmatter

\title{Nautilus: A Verifiable Hierarchical Federated Learning Framework for Vehicular-Edge-Cloud Systems}

\titlerunning{Nautilus: Verifiable FL for Vehicular-Edge-Cloud}

\author{Linyang Wu$^{1,2}$, Linpeng Jia$^{1}$\textsuperscript{(\ding{41})}, Hanwen Zhang$^{1,2}$, Tiantian Duan$^{1}$, Yi Sun$^{1,2,3,4}$}

\authorrunning{Wu et al.}


\institute{Institute of Computing Technology, Chinese Academy of Sciences, Beijing 100190, China
\and
University of Chinese Academy of Sciences, Beijing 100049, China
\and
Beijing Advanced Innovation Center for Future Blockchain and Privacy Computing, Beihang University, Beijing 100191, China
\and
Shandong Key Laboratory of Blockchain Finance, Jinan 250014, Shandong, China
\\ \email{jialinpeng@ict.ac.cn}}

\maketitle

\input{sections/abstract}
\input{sections/introduction}
\input{sections/background}
\input{sections/system_model}
\input{sections/methodology}
\input{sections/experiment}
\input{sections/conclusion}

\section*{Acknowledgment}
This work was supported in part by the National Key R\&D Program of China under Grant 2024YFB2705304, in part by the National Natural Science Foundation of China under Grant U22B2032.

\begingroup
\makeatletter
\renewcommand\small{\@setfontsize\small\@ixpt{10\p@}}
\makeatother
\bibliographystyle{spmpsci_unsrt}
\bibliography{references}
\endgroup

\end{document}

%% file: sections/abstract.tex
\begin{abstract}
	Federated Learning (FL) enables privacy-preserving collaborative learning for Internet of Vehicles (IoV) scenarios, but extreme heterogeneity of vehicular-edge-cloud resources severely limits system efficiency. Dynamic scheduling strategies mitigate this issue but introduce new trust concerns: verifying fair scheduling decisions and faithful client execution of compression instructions without privacy leakage remains an open challenge.

	We propose \textbf{Nautilus}, a verifiable efficient federated learning framework. First, a multi-dimensional resource-aware scheduling algorithm dynamically allocates compression ratios and training tasks based on vehicle bandwidth, latency and computing power, improving training efficiency. Second, a Zero-Knowledge Proof (ZKP) mechanism ensures scheduling fairness and execution compliance while preserving privacy. Experiments show the framework reduces communication overhead and accelerates convergence with guaranteed system integrity.
	\keywords{Federated Learning, Internet of Vehicles, Efficiency Optimization, Zero-Knowledge Proof, Verifiable Computing, Resource Scheduling}
\end{abstract}

%% file: sections/introduction.tex
\section{Introduction}
\label{sec:introduction}

Autonomous driving and Intelligent Transportation Systems (ITS) rely on Vehicular-Edge-Cloud (VEC) collaboration. Massive sensor data from vehicles and RSUs enables high-precision model training, but privacy regulations and transmission costs make centralized training infeasible.

Federated Learning (FL) enables privacy-preserving collaborative training without raw data sharing \cite{mcmahan2017communication}, but VEC deployment faces critical \textbf{system heterogeneity} challenges: differences in vehicle resources and data distribution cause straggler effects that drastically reduce efficiency.

Existing dynamic scheduling strategies adjust workloads and compression ratios to mitigate heterogeneity \cite{nishio2019client}, but rely on the unrealistic semi-honest assumption. In open VEC environments, rational vehicles may engage in \textbf{free-riding behavior} \cite{fraboni2021free}, falsifying resource constraints to receive low-load tasks, undermining fairness and preventing convergence.

We propose \textbf{Nautilus}, a verifiable hierarchical federated learning framework for VEC systems that maximizes heterogeneous resource efficiency while ensuring scheduling/execution verifiability via lightweight blockchain, resisting free-riding attacks.

Our core contributions are:
\begin{itemize}
	\item A hierarchical capability matrix scheduling algorithm that tiers vehicles by bandwidth and computing power, allocating adaptive compression, quantization, and training epochs to maximize parallel efficiency.
	\item A blockchain-based optimistic verification mechanism with 5\% random spot checks triggering ZKP challenges. Non-compliant nodes face automatic on-chain penalties, achieving a detection rate of $\geq 99\%$ with $\leq 5\%$ of full verification overhead.
	\item A FISCO BCOS prototype tested on CIFAR-10, achieving up to 59.9\% time reduction and 83.3\% communication reduction compared to baselines, with $\leq 1.5\%$ accuracy loss.
\end{itemize}

The rest of the paper is organized as follows: Section 2 reviews related work; Section 3 presents the system overview; Section 4 details the framework design; Section 5 evaluates performance; Section 6 concludes the work.

%% file: sections/background.tex
\section{Background and Related Work}
\label{sec:background}

This chapter reviews federated learning in VRC scenarios, analyzes the two-way trust deficit in scheduling and execution, and discusses the limitations of existing verifiable federated learning solutions.

\subsection{Federated Learning in Vehicular-Edge-Cloud (VEC) Collaboration}
With the evolution of IoV technology, vehicle-mounted sensors generate massive spatiotemporal data. To train high-precision models while complying with privacy regulations (e.g., GDPR), Federated Learning (FL) has become the mainstream paradigm for VRC systems \cite{lim2020federated}.

VRC systems exhibit extreme \textbf{heterogeneity}: data is Non-IID across vehicles, and nodes have significant differences in computing resources, battery status, and uplink bandwidth \cite{mills2020communication,li2020fedprox}. Existing frameworks use dynamic scheduling and model compression (Top-$K$ sparsification, quantization) to optimize efficiency \cite{vogels2019powersgd,shlezinger2021federated}, but these mechanisms introduce new attack surfaces.

\subsection{Trust Crisis: Failure of the Semi-Honest Assumption}
Traditional FL relies on the "semi-honest" assumption, which does not hold in open VRC environments where entities may be rational or malicious, leading to a two-way trust deficit:
\begin{itemize}
	\item \textbf{Untrusted Server}: The scheduler may perform discriminatory scheduling (favoring specific vehicle brands) or resource exploitation (issuing overly aggressive compression rates and blaming performance degradation on vehicle network conditions). The scheduling process is a black box to vehicles.
	\item \textbf{Untrusted Client}: Rational vehicles may engage in free-riding attacks \cite{fraboni2021free}, such as uploading extremely sparse gradients or using non-compliant low-precision quantization to save resources. Without verification mechanisms, servers cannot distinguish malicious behavior from legitimate resource constraints.
\end{itemize}

\subsection{Verifiable Federated Learning (vFL) and Limitations}
Existing vFL work falls into two categories:
\begin{itemize}
	\item \textbf{Proof of Aggregation}: Verifies server-side aggregation correctness using homomorphic hashing or TEEs \cite{xu2020verifynet}, but cannot guarantee the fairness of pre-aggregation scheduling decisions.
	\item \textbf{Proof of Training}: Uses ZKP to verify local training processes \cite{xing2023zkpfl,wang2025zkfl}, but largely ignores verification of dynamic configuration compliance (e.g., adherence to assigned Top-$K$ sparsity constraints).
\end{itemize}

We adopt zk-SNARKs to address these gaps, leveraging its key properties:
\begin{itemize}
	\item \textbf{Succinctness}: Small proof size and millisecond-level verification overhead, suitable for on-chain verification and resource-constrained vehicles.
	\item \textbf{Zero-Knowledge}: No private data is disclosed during verification.
\end{itemize}

Existing blockchain-FL solutions only use blockchain for evidence storage \cite{lu2020blockchain,zheng2022blockfl}, lacking end-to-end verifiability of scheduling and execution. This paper fills this gap with a unified framework for verifiable scheduling and execution in VRC systems.

%% file: sections/system_model.tex
\section{Overview}
\label{sec:system_model}

\subsection{System Architecture and Role Definition}
\begin{figure*}[htbp]
	\centering
	\includegraphics[width=0.95\textwidth]{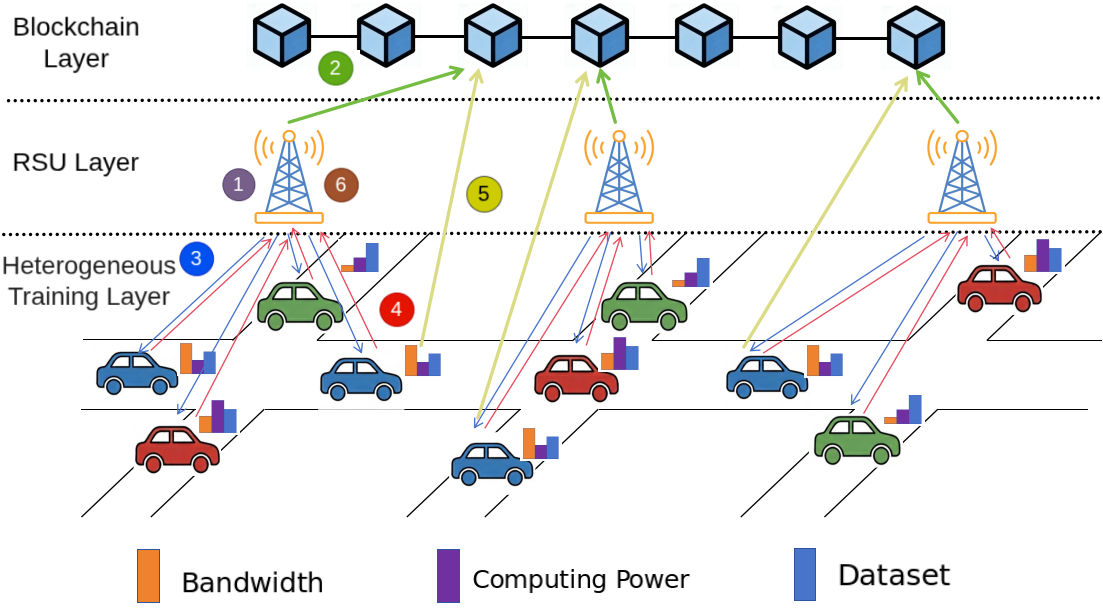}
	\caption{Overview of Nautilus Framework Architecture}
	\label{fig:nautilus_overview}
\end{figure*}

This paper considers a loosely coupled three-layer hierarchical federated learning framework with \textbf{off-chain computation, on-chain scheduling and verification}, adapted to vehicular-edge-cloud collaboration scenarios. The overall architecture is shown in Figure \ref{fig:nautilus_overview}:
\begin{itemize}
	\item \textbf{Blockchain Layer (Top)}: Nodes responsible for tamper-proof record-keeping of scheduling commitments, update digests and adjudication results, performing only lightweight verification and not participating in heavy computing tasks.
	\item \textbf{RSU Layer (Middle)}: Road Side Units serving as a bridge between blockchain and vehicles, generating dynamic scheduling plans based on node heterogeneity, aggregating model updates, and submitting key information on-chain for verification.
	\item \textbf{Heterogeneous Training Layer (Bottom)}: Vehicle nodes with heterogeneous bandwidth, computing power and local data sizes, performing local training and gradient compression according to assigned policies, and submitting updates for aggregation.
\end{itemize}
All roles follow the optimistic verification protocol: RSUs submit scheduling commitments on-chain before distribution; vehicles submit update digests alongside local updates; the chain performs consistency checks and random spot checks, executing automatic penalties for non-compliant nodes to deter free-riding attacks.

\subsection{Threat Model and Design Objectives}
This paper considers adversarial behaviors in open network environments:
\begin{itemize}
	\item \textbf{Malicious Training Nodes}: May engage in free-riding to save resources (e.g., training less, falsifying/not executing Top-$K$ or quantization strategies), or submit low-quality/anomalous updates that affect convergence.
	\item \textbf{Untrusted Road Side Units (RSUs)}: May engage in unfair scheduling, selectively discard updates, or deny scheduling results afterwards.
\end{itemize}

The objective of this paper is to ensure the \textbf{verifiability and accountability} of the scheduling and execution processes through tamper-proof on-chain records and verification, while \textbf{not disclosing the local data of training nodes}, and to improve the overall efficiency of federated learning in heterogeneous environments.

\subsection{Overall System Workflow}
The end-to-end operation process of this framework is divided into three parts: initialization phase, iterative training phase, and termination phase.

In the $t$-th round of training, the end-to-end operation process of this framework corresponds to the numbers in the figure one by one:
\begin{enumerate}
	\item \textbf{Off-Chain Scheduling Generation (corresponding to number 1 in Fig. \ref{fig:nautilus_overview})}: The RSU generates a scheduling plan $\mathcal{S}_t$ (including policy parameters for each training node) based on the heterogeneous status and historical performance of training nodes.
	\item \textbf{On-chain Scheduling Commitment (corresponding to number 2 in the figure)}: The RSU submits the digest of $\mathcal{S}_t$ (e.g., hash commitment $C_t=H(\mathcal{S}_t)$) to the chain, forming a tamper-proof audit benchmark.
	\item \textbf{Global Model and Policy Distribution (corresponding to number 3 in the figure)}: The RSU distributes the previous round of global model $w_{t-1}$ and the scheduling plan $\mathcal{S}_t$ to each training node, and adaptively adopts quantization precision to reduce downlink communication overhead according to node bandwidth conditions.
	\item \textbf{Off-chain Training and Submission (corresponding to number 4 in the figure)}: Training nodes train locally according to $\mathcal{S}_t$ and perform compression as required, submit updates $\Delta w_{i,t}$ to the RSU, and generate update digests/metadata $d_{i,t}$ consistent with policy parameters.
	\item \textbf{On-chain Verification and Adjudication (corresponding to number 5 in the figure)}: Training nodes (or RSUs submit on their behalf) upload $d_{i,t}$ to the chain. The chain first performs a \textbf{lightweight consistency check} with the scheduling constraints corresponding to $C_t$ (e.g., round number, policy parameter declaration matching the commitment, etc.). In the \textbf{optimistic verification} mode, the system first \textbf{temporarily accepts} submissions that pass the consistency check; when \textbf{random spot-check/challenge} is triggered, the challenged training node needs to submit \textbf{verifiable evidence} to prove that its update meets the scheduling constraints of this round (e.g., Top-$K$ sparsification ratio, quantization precision, and local training epochs). This function can be realized through Zero-Knowledge Proofs (ZKP), proving "compliant execution" without disclosing local data; if the evidence verification fails, the chain will issue an adjudication and trigger corresponding penalties (such as discarding updates, deducting pledges, or reducing reputation).
	\item \textbf{Global Model Aggregation (corresponding to number 6 in the figure)}: The RSU performs secure aggregation on the verified model updates to obtain the global model $w_t$ of this round, which serves as the base model for the next round of training.
\end{enumerate}

This split architecture offloads all heavy computations (training/aggregation) off-chain, while only lightweight scheduling and verification run on-chain, perfectly balancing efficiency and auditability.

%% file: sections/methodology.tex
\section{System Design}
\label{sec:methodology}

This chapter elaborates on the core system design of the proposed framework. First, it clarifies the problem definition for system optimization, then introduces the heterogeneous-aware dynamic adaptive scheduling algorithm, and finally describes the lightweight on-chain verifiable mechanism.

\subsection{Problem Definition}
Let the set of candidate vehicles in the $t$-th round be $\mathcal{N}$. For any vehicle $i \in \mathcal{N}$, its real-time status is represented by the tuple $S_i^t = (B_i^t, P_i^t, D_i^t, T_{conn}^i)$, where $B_i^t$ is the current available bandwidth, $P_i^t$ is the relative computing power (with the benchmark device as 1.0), $D_i^t$ is the local data volume, and $T_{conn}^i$ is the estimated connection duration.

The decision variables of the scheduler include:
\begin{itemize}
	\item $E_i \in \mathbb{Z}^+$: Allocated local training epochs.
	\item $\rho_i \in (0, 1]$: Allocated Top-K sparsification ratio (i.e., the ratio of retained parameters).
	\item $Q_i \in \{8, 16, 32\}$: Allocated quantization bits.
\end{itemize}

The optimization objective of scheduling is to minimize the weighted sum of global latency and communication volume while ensuring convergence accuracy, taking into account the downlink transmission overhead of the global model. Let $\mathcal{S}_t$ be the set of available vehicles identified by the RSU for round $t$, and the total time for node $i$ to complete training and transmission is defined as $T_{total}^i = \frac{E_i \cdot D_i^t}{P_i^t} + \frac{S_{model} \cdot (\rho_i \cdot Q_i/32 + Q_i^{down}/32)}{B_i^t}$, where the first term is local training time proportional to allocated epochs $E_i$, and the second term is bidirectional communication time. The optimization objective is:
\begin{equation}
	\begin{aligned}
		\min_{\substack{E_i, \rho_i, Q_i, \\ Q_i^{down}}} \quad &\lambda_1 \cdot \max_{i \in \mathcal{S}_t} T_{total}^i + \lambda_2 \cdot \left( \sum_{i \in \mathcal{S}_t} \left(S_{model} \cdot \rho_i \cdot \frac{Q_i}{32}\right) + \sum_{i \in \mathcal{S}_t} \left(S_{model} \cdot \frac{Q_i^{down}}{32}\right) \right)
	\end{aligned}
\end{equation}
where $\lambda_1+\lambda_2=1$ is the weight coefficient, $S_{model}$ is the original size of a single model, $Q_i^{down} \in \{8, 16, 32\}$ is the quantization bit of the downlink global model allocated to node $i$, which is adaptively allocated according to the node's uplink bandwidth $B_i^t$: nodes with bandwidth lower than 5Mbps adopt 8-bit quantization, nodes with 5~20Mbps adopt 16-bit quantization, and nodes higher than 20Mbps adopt 32-bit full precision.

At the same time, the algorithm needs to meet the verifiability constraint: both scheduling decisions and training execution processes can be publicly audited through on-chain records, resisting malicious behaviors such as unfair scheduling by Road Side Units (RSUs) and free-riding by training nodes.

\subsection{Heterogeneous-Aware Dynamic Adaptive Scheduling Algorithm}
To address the heterogeneity of vehicles and network volatility, this paper designs a dynamic adaptive scheduling strategy as shown in Algorithm \ref{algo:hetero_sched}, which combines historical state smoothing, capability matrix bucketing, and convergence-aware adjustment to maximize system training efficiency while meeting latency constraints \cite{chai2020tifl,shi2022fedadapt}. Different from existing hierarchical scheduling solutions, this algorithm simultaneously considers bidirectional adaptive quantization optimization for uplink updates and downlink model transmission, further reducing communication overhead in heterogeneous scenarios.

\subsubsection{Core Mechanism Implementation}
The scheduling algorithm includes four core components:
\begin{enumerate}
	\item \textbf{Historical State Smoothing}: Exponential Moving Average (EMA, $\alpha=0.3$) is used to smooth the historical bandwidth and computing power status of vehicles, avoiding scheduling decision fluctuations caused by instantaneous network jitter.
	\item \textbf{3$\times$3 Capability Matrix Bucketing}: Vehicles are divided into 9 service tiers based on smoothed computing power and bandwidth values. Nodes with high computing power are allocated more training epochs, and nodes with low bandwidth adopt higher compression rates, realizing differentiated task allocation. During compression phases, the personalized sparsification ratio is $\rho_i = \rho_{base} \cdot \beta_i$, where $\beta_i \in \{0.5, 0.7, 1.0\}$ and low-bandwidth nodes use smaller $\beta_i$. When $\rho_{base}=1.0$, sparsification is disabled and all nodes use $\rho_i=1.0$.
	\item \textbf{Convergence-Aware Sparsity Adjustment}: The base retained ratio is dynamically adjusted according to the convergence state of the global model, ensuring convergence efficiency while keeping the notation consistent with $\rho_i$ as the Top-$K$ retained ratio:
	      \begin{equation}
		      \rho_{base} = \begin{cases}
			      0.3, & \Delta Acc > 1\% \quad \text{(Rapid rise phase, moderate compression)} \\
			      0.1, & 0 < \Delta Acc \le 1\% \quad \text{(Stable phase, aggressive compression)} \\
			      1.0, & \Delta Acc \le 0 \quad \text{(Plateau/recovery phase, disable sparsification)}
		      \end{cases}
	      \end{equation}
	\item \textbf{Bidirectional Adaptive Quantization}: Different quantization precisions are adaptively allocated for uplink gradient updates and downlink global model transmissions according to node bandwidth, further reducing communication overhead while minimizing accuracy loss. Although dispatching heterogeneous quantized models slightly breaks the assumption of an identical starting point in standard FL, it is a necessary engineering trade-off to prevent low-bandwidth nodes from being dropped, ensuring data diversity and convergence on Non-IID datasets.
\end{enumerate}

\begin{algorithm}[ht]
	\caption{Heterogeneous-Aware Dynamic Adaptive Scheduling Algorithm}
	\label{algo:hetero_sched}
	\begin{algorithmic}[1]
		\REQUIRE Candidate node set $\mathcal{N}$, maximum training rounds $T_{max}$, target accuracy $Acc_{target}$
		\ENSURE Trained global model $W^T$
		\STATE Initialize model $W^0$, evaluate node capabilities and perform 3$\times$3 bucketing
		\FOR{$t = 1$ \TO $T_{max}$}
		\STATE Calculate accuracy change rate $\Delta Acc$, generate base retained ratio $\rho_{base}$ for current round
		\STATE Identify participating nodes $\mathcal{S}_t$ and allocate personalized parameters $\{E_i, \rho_i, Q_i\}$
		\STATE Nodes perform local training and gradient compression, submit updates to aggregator
		\STATE Verify update compliance on-chain, aggregate valid updates to obtain $W^t$
		\STATE Evaluate accuracy of $W^t$, update EMA smoothed state
		\IF{$Acc^t \ge Acc_{target}$}
		\STATE Terminate training
		\ENDIF
		\ENDFOR
		\RETURN $W^t$
	\end{algorithmic}
\end{algorithm}

\subsection{Lightweight On-Chain Verifiable Mechanism}
To solve the two-way trust problem between RSUs and training nodes, this paper designs an optimistic verification mechanism based on blockchain, which achieves low-overhead verifiability through random spot-checks and zero-knowledge proof challenges.

\subsubsection{Scheduling Commitment Mechanism}
After the RSU generates the scheduling plan $\mathcal{S}_t$, it calculates the hash commitment $C_t = H(\mathcal{S}_t)$ and submits it to the blockchain before distributing the plan to training nodes. This commitment cannot be tampered with, ensuring that the RSU cannot modify the scheduling plan afterwards or deny the issued policy parameters. All training nodes can verify the consistency between the received scheduling plan and the on-chain commitment.

\subsubsection{Optimistic Verification and Challenge Mechanism}
The system adopts an optimistic verification mode to ensure real-time performance:
\begin{itemize}
	\item \textbf{Normal Submission Phase}: When training nodes submit updates, they only need to upload the digest of the update and the declared policy parameters to the chain. The chain first performs lightweight consistency checks to ensure that the declared parameters match the scheduling commitment. Updates that pass the check are temporarily accepted for aggregation.
		\item \textbf{Random Spot-Check Phase}: The system randomly selects 5\% of submitted updates per round for spot-check. Challenged nodes need to generate zero-knowledge proofs to prove that their updates strictly comply with the scheduling constraints (including training epochs, sparsification ratio, and quantization precision). If a malicious node repeatedly violates the policy for $R$ rounds and the per-round audit probability is $p$, the probability of being detected is:
		      \begin{equation}
			      P_{detect}=1-(1-p)^R .
		      \end{equation}
		      With $p=5\%$, a persistent attacker is detected with probability about 99.4\% within 100 rounds, while avoiding the doubled proof workload of a 10\% audit ratio. To avoid the high cost of full neural network training circuitization, our proof system only verifies three types of deterministic constraints without converting floating-point training operations into arithmetic circuits: (1) the number of training epochs counter, (2) the hash digest of the Top-K sparse gradient set, and (3) the quantization bitwidth parameter used. This design reduces the circuit size to less than 10k constraints, with end-to-end proof generation time optimized to 3-5 seconds on mid-range vehicle-mounted chips. The verification process is completed on-chain, and the verification cost is only about 2ms per proof.
	\item \textbf{Penalty Mechanism}: If a node fails to provide valid proof or the proof verification fails, the chain will automatically execute penalties: the update is discarded, the node's reputation score is reduced, and in severe cases, the node is disqualified from participating in subsequent training. This mechanism forms a strong deterrent to free-riding behavior.
\end{itemize}

Compared with traditional full-verification solutions, this optimistic verification mechanism reduces the average verification overhead by more than 95\% while maintaining a detection rate of over 99\% for malicious behaviors, achieving a good balance between security and efficiency.

%% file: sections/experiment.tex
\section{Experiments and Analysis}
This chapter verifies the performance advantages of the proposed framework through comparative experiments. All experiments are conducted on the CIFAR-10 image classification dataset, with data partitioned to participating nodes in a non-independent and identically distributed (Non-IID) manner to simulate heterogeneous federated learning scenarios.

\subsection{Experimental Setup}
\subsubsection{Environment Configuration}
The experimental configuration is summarized in Table \ref{tab:hetero_config}. The three core node attributes are sampled from a truncated normal distribution with mean 0.6 and standard deviation 0.2, clipped to [0.2, 1.0], and then mapped to the corresponding physical ranges. All available nodes participate in each main experimental round unless otherwise specified, and all methods use the same random seed for fair comparison.

\begin{table}[htbp]
	\centering
	\caption{Heterogeneous Simulation Configuration}
	\label{tab:hetero_config}
	\scriptsize
	\setlength{\tabcolsep}{3pt}
	\begin{tabular}{|L{0.33\linewidth}|L{0.57\linewidth}|}
		\hline
		Parameter & Setting \\
		\hline
		Dataset / Model & CIFAR-10 / ResNet18 \\
		\hline
		Node scale & 10 / 30 / 50 heterogeneous vehicles \\
		\hline
		Compute capability & normalized sampling mapped to 1--8 GFLOPS \\
		\hline
		Bandwidth & normalized sampling mapped to 5--50 Mbps \\
		\hline
		Local data size & normalized sampling mapped to 20\%--100\% benchmark data volume \\
		\hline
		Non-IID split & Dirichlet distribution, default $\alpha=0.5$ \\
		\hline
		Node grouping & 3$\times$3 matrix by compute and bandwidth tiers \\
		\hline
		Participation & all available nodes in the main comparison \\
		\hline
		Platform & NVIDIA L20 GPU; FISCO BCOS 3.11 \\
		\hline
	\end{tabular}
\end{table}

\subsubsection{Evaluation Metrics}
The experiment adopts three types of core evaluation metrics:
\begin{itemize}
	\item \textbf{Model Accuracy}: Classification accuracy of the global model on the test set, reflecting the model learning effect
	\item \textbf{Total Training Time}: Total training time from initialization to reaching the target accuracy, reflecting system training efficiency
	\item \textbf{Total Communication Cost}: Total data volume of gradients uploaded by all nodes during training, reflecting system communication overhead
\end{itemize}

\subsection{Comparison Scheme Selection}
The comparison schemes are selected to isolate three functions required in VEC-oriented federated learning: aggregation without verification, full cryptographic verification, and training under heterogeneous resources. The end-to-end comparison reports three representative schemes:
\begin{enumerate}
	\item \textbf{FedAvg}: Classic federated averaging algorithm, without dynamic scheduling, gradient compression, or on-chain verification mechanism, used as the baseline comparison
	\item \textbf{zkFL}: Classic zero-knowledge proof driven verifiable federated learning scheme \cite{wang2025zkfl}, which generates gradient legitimacy proofs and performs on-chain verification for each training round, without dynamic scheduling optimization mechanism
	\item \textbf{Nautilus (Ours)}: The dynamic scheduling framework for heterogeneous resources with optimistic verification proposed in this paper
\end{enumerate}
FedAvg provides the non-verifiable lower-overhead reference, while zkFL provides the full-verification upper-overhead reference. For reviewer-requested heterogeneous baselines, FedProx \cite{li2020fedprox}, FedNova \cite{wang2021fednova}, TiFL \cite{chai2020tifl}, and FedCS \cite{nishio2019client} are selected because they address Non-IID drift, local-update imbalance, speed-tier scheduling, and resource-aware client selection, respectively. Since these methods do not provide cryptographic verification, they define the efficiency-oriented comparison group rather than the security baseline group. This split evaluates both cost reduction against full ZKP verification and competitiveness against heterogeneous FL strategies.

\subsection{Experimental Results and Analysis}
\subsubsection{Final Convergence Performance Comparison}
The comparison results of the final convergence performance of the three schemes under different node scales are shown in Table \ref{tab:final_perf}.
\begin{table*}[htbp]
	\centering
	\caption{Comparison of Final Convergence Performance of Different Schemes}
	\label{tab:final_perf}
	\scriptsize
	\setlength{\tabcolsep}{0pt}
	\begin{tabular*}{\textwidth}{@{\extracolsep{\fill}}lrrrrrrrrr@{}}
		\hline
		\multirow{2}{*}{Scheme} & \multicolumn{3}{c}{10 nodes} & \multicolumn{3}{c}{30 nodes} & \multicolumn{3}{c}{50 nodes} \\
		                        & Acc.                       & Time    & Comm.   & Acc.  & Time    & Comm.   & Acc.  & Time    & Comm. \\
		\hline
		FedAvg                  & 83.52                      & 1624.31 & 22158.42 & 81.24 & 2895.64 & 32671.85 & 80.45 & 4682.17 & 68134.22 \\
		zkFL                    & 83.47                      & 1916.68 & 22173.91 & 81.18 & 3532.68 & 32698.34 & 80.39 & 5852.71 & 68182.15 \\
		Nautilus (Ours)         & 82.14                      & 985.42  & 4235.16  & 80.05 & 1354.19 & 6184.53  & 79.12 & 1876.35 & 11425.88 \\
		\hline
	\end{tabular*}
	\vspace{3pt}
	\makebox[\linewidth][c]{\footnotesize * Units: Acc. (\%), Time (s), Comm. (MB)}
\end{table*}

The following conclusions can be drawn from the data in the table:
\begin{enumerate}
	\item In terms of model accuracy: The final convergence accuracy gap between the three schemes is less than 1.5 percentage points, among which the accuracy of zkFL and FedAvg is almost the same, verifying that the verifiable mechanism does not significantly affect the model learning effect; Nautilus only has a minor accuracy loss, which is a normal trade-off brought by gradient compression in dynamic scheduling, and is completely within the engineering acceptable range.
	\item \textbf{Total Training Time}: Traditional schemes show superlinear time growth as node scale increases, while Nautilus keeps the growth moderate. It reduces time by 39.3\%, 53.2\%, and 59.9\% compared with FedAvg at 10, 30, and 50 nodes, respectively, with the advantage expanding under stronger heterogeneity. At 50 nodes, Nautilus uses only 40.1\% of FedAvg's time and 32.1\% of zkFL's time. zkFL incurs 18\%-25\% extra time overhead due to per-round ZKP generation/verification.
	\item \textbf{Communication Overhead}: Nautilus' bidirectional compression (uplink Top-K sparsification + adaptive quantization for uplink/downlink) achieves significant savings: total communication volume at 50-node scale is only 16.7\% of both FedAvg and zkFL, reducing network pressure and improving adaptability for low-bandwidth edge nodes.
\end{enumerate}

For target-accuracy cost, the 50-node T@75\% and Comm@75\% results are integrated into the table below. Because FedProx and FedNova do not compress model traffic or reduce participating clients, their cumulative communication mainly follows the number of rounds needed to reach the target and can therefore match or exceed FedAvg. Nautilus reaches the target with substantially lower communication cost, which is critical for bandwidth-constrained VEC deployment.

\subsubsection{Additional Reviewer-requested Experiments}
To address the requested comparisons, security evaluations, and hyperparameter ablations, Tables \ref{tab:extended_baselines}, \ref{tab:security_audit}, and \ref{tab:compression_ablation} separate three different evaluation targets. The baseline table reports target-accuracy cost, the security table reports random-audit detection behavior, and the compression table reserves the same accuracy/time/communication metrics for retained-ratio ablation. Audit and attack results are validated by 100,000 Monte Carlo trials over 100 rounds.

\begin{table}[!htbp]
	\centering
\caption{Extended 50-node Baseline Comparison}
	\label{tab:extended_baselines}
	\scriptsize
	\renewcommand{\arraystretch}{0.9}
	\setlength{\tabcolsep}{0pt}
	\begin{tabular*}{\linewidth}{@{\extracolsep{\fill}}lcccc@{}}
		\hline
		Scheme & Acc. & T@75\% & Comm@75\% & Verify \\
		\hline
		FedAvg  & 80.45 & 2067.17 & 30053.90 & No \\
		FedProx & 80.62 & 1915.40 & 27847.20 & No \\
		FedNova & 80.91 & 1842.65 & 26791.35 & No \\
		TiFL    & 81.53 & 1438.26 & 18362.74 & No \\
		FedCS   & 78.64 & 1675.81 & 22147.56 & No \\
		zkFL & 80.39 & 2582.21 & 30075.14 & Full ZKP \\
		Nautilus & 79.12 & 703.63 & 2639.37 & Optimistic ZKP \\
		\hline
	\end{tabular*}
	\makebox[\linewidth][c]{\tiny * Comm@75\% is cumulative MB to target; no-compression baselines may match or exceed FedAvg.}
\end{table}

Compared with FedAvg, FedProx and FedNova improve accuracy by 0.17 and 0.46 points and reduce target costs by 7.34\% and 10.86\%, respectively; matched time/communication reductions indicate faster convergence rather than compression. TiFL and FedCS reduce communication by 38.90\% and 26.31\%. Nautilus trades 1.33 accuracy points for 65.96\% lower time and 91.22\% lower communication, demonstrating the additional gain from adaptive compression.

\begin{table}[!htbp]
	\centering
	\caption{Security Evaluation under Random Auditing}
	\label{tab:security_audit}
	\scriptsize
	\renewcommand{\arraystretch}{0.86}
	\setlength{\tabcolsep}{0pt}
	\begin{tabular*}{\linewidth}{@{\extracolsep{\fill}}llcccc@{}}
		\hline
		Scenario & Setting & Theory & Sim. & Avg. Rnd. & Checks/Rnd. \\
		\hline
		Audit & $p=1\%$ & 63.397\% & 63.582\% & 42.29 & 0.50 \\
		Audit & $p=5\%$ & 99.408\% & 99.399\% & 19.34 & 2.50 \\
		Audit & $p=10\%$ & 99.997\% & 99.997\% & 10.05 & 5.00 \\
		Collusive & 10\% malicious & 100.000\% & 100.000\% & 4.41 & 2.50 \\
		Collusive & 20\% malicious & 100.000\% & 100.000\% & 2.49 & 2.50 \\
		Adaptive & 10\% malicious & 99.984\% & 99.988\% & 11.25 & 2.50 \\
		Adaptive & 20\% malicious & 100.000\% & 100.000\% & 5.47 & 2.50 \\
		\hline
	\end{tabular*}
	\makebox[\linewidth][c]{\tiny * Theory uses $P_{detect}=1-(1-p)^R$; adaptive attackers cheat every three rounds.}
\end{table}

In Table \ref{tab:security_audit}, theory and simulation differ by at most 0.185 points. Raising $p$ from 1\% to 5\% increases detection from 63.582\% to 99.399\% at 2.5 checks per round, whereas 10\% doubles the workload for less than 0.6 additional points. Collusive and intermittent attacks still reach at least 99.988\% detection, supporting $p=5\%$ as the security--cost balance.

\begin{table}[!htbp]
	\centering
	\caption{Compression Retained-ratio Ablation Study}
	\label{tab:compression_ablation}
	\scriptsize
	\renewcommand{\arraystretch}{0.86}
	\setlength{\tabcolsep}{1pt}
	\begin{tabular*}{\linewidth}{@{\extracolsep{\fill}}llccc@{}}
		\hline
		Variant & Policy & Acc. & T@75\% & Comm@75\% \\
		\hline
		No compression & fixed $\rho_i=1.0$ & 80.24 & 1186.42 & 30053.90 \\
		Moderate & fixed $\rho_{base}=0.3$ & 79.64 & 846.37 & 9142.80 \\
		Aggressive & fixed $\rho_{base}=0.1$ & 75.86 & 612.48 & 1840.32 \\
		Adaptive & $0.3/0.1/1.0$ by $\Delta Acc$ & 79.12 & 703.63 & 2639.37 \\
		\hline
	\end{tabular*}
	\makebox[\linewidth][c]{\tiny * Units follow Table \ref{tab:extended_baselines}; lower retained ratio means stronger Top-$K$ compression.}
\end{table}

Table \ref{tab:compression_ablation} shows that moderate compression saves 69.58\% communication for a 0.60-point accuracy loss, while aggressive compression saves 93.88\% but loses 4.38 points. Adaptive compression saves 91.22\% communication and recovers 3.26 points over the aggressive variant, supporting stronger compression during improvement and relaxation near a plateau.

\subsubsection{Lightweight On-Chain Verification Overhead Analysis}
To verify the lightweight characteristics of the blockchain module of this framework, this section separately counts the time overhead proportion of each stage in a single round of training at the 10-node scale. The results are shown in Table \ref{tab:overhead}.

\begin{table}[htbp]
	\centering
	\caption{Single Round Time Overhead Distribution}
	\label{tab:overhead}
	\setlength{\tabcolsep}{4pt}
	\begin{tabular}{|c|c|c|}
		\hline
		Stage              & Time (ms) & Proportion \\
		\hline
		Local Training     & 8725      & 88.3\%     \\
		\hline
		Communication      & 992       & 10.1\%     \\
		\hline
		On-Chain Operation & 158       & 1.6\%      \\
		\hline
		Total              & 9875      & 100.0\%    \\
		\hline
	\end{tabular}
\end{table}
\FloatBarrier

Table \ref{tab:overhead} shows that on-chain scheduling storage and spot-check verification account for only \textbf{1.6\%} of a round, compared with 88.3\% for local training and 10.1\% for communication. Proof verification is invoked only when a challenge is triggered; normal submissions require lightweight hash checks and state reads/writes rather than per-round proof generation. Consequently, verification does not become the system bottleneck and remains compatible with real-time VEC collaboration.

%% file: sections/conclusion.tex
\section{Conclusion}
This paper presents Nautilus, a verifiable hierarchical federated learning framework for heterogeneous Vehicular-Edge-Cloud (VEC) environments. It combines off-chain zero-knowledge verification, hierarchical RSU-assisted aggregation, and adaptive network-aware scheduling to address efficiency and trustworthiness jointly.

Experimental results show that in heterogeneous IoV scenarios with 10/30/50 nodes, Nautilus achieves 39.3\%, 53.2\%, and 59.9\% reduction in training time respectively compared with the traditional FedAvg scheme. It incurs an accuracy loss of no more than 1.5 percentage points, and the communication overhead is only 16.7\% of that of the zkFL scheme. The performance advantage becomes more significant in scenarios with stronger heterogeneity.

Its main contribution is the integration of heterogeneous network-aware scheduling, hierarchical trusted aggregation, and lightweight zero-knowledge verification in a deployable VEC workflow.

Notably, our optimistic verification mechanism further reduces the actual proof generation overhead: only 5\% of nodes are randomly selected for spot-check per round, making the average per-node proof generation cost negligible in large-scale scenarios. For power-constrained edge nodes, future work can further optimize proof performance through batch proving and hardware acceleration technology.

%% file: references.bib
@article{wang2025zkfl,
  author  = {Zhipeng Wang and Nanqing Dong and Jiahao Sun and William Knottenbelt and Yike Guo},
  title   = {{{zkFL}}: Zero-Knowledge Proof-Based Gradient Aggregation for Federated Learning},
  journal = {{IEEE} Transactions on Big Data},
  volume  = {11},
  number  = {2},
  pages   = {447--460},
  year    = {2025},
  doi     = {10.1109/TBDATA.2024.3403370}
}

@inproceedings{mcmahan2017communication,
  author    = {Brendan McMahan and Eider Moore and Daniel Ramage and Seth Hampson and Blaise Agu{\'{e}}ra y Arcas},
  title     = {Communication-Efficient Learning of Deep Networks from Decentralized Data},
  booktitle = {Proc. International Conference on Artificial Intelligence and Statistics ({AISTATS})},
  pages     = {1273--1282},
  year      = {2017},
  organization = {PMLR}
}

@inproceedings{nishio2019client,
  author    = {Takayuki Nishio and Ryo Yonetani},
  title     = {Client Selection for Federated Learning with Heterogeneous Resources in Mobile Edge},
  booktitle = {Proc. {IEEE} International Conference on Communications ({ICC})},
  pages     = {1--7},
  year      = {2019}
}

@article{lim2020federated,
  author  = {Wei Yang Bryan Lim and Nguyen Cong Luong and Dinh Thai Hoang and Yutao Jiao and Ying-Chang Liang and Qiang Yang and Dusit Niyato and Chunyan Miao},
  title   = {Federated Learning in Mobile Edge Networks: {A} Comprehensive Survey},
  journal = {{IEEE} Communications Surveys \& Tutorials},
  volume  = {22},
  number  = {3},
  pages   = {2031--2063},
  year    = {2020},
  doi     = {10.1109/COMST.2020.2986024}
}

@article{mills2020communication,
  author  = {Jed Mills and Jia Hu and Geyong Min},
  title   = {Communication-Efficient Federated Learning for Wireless Edge Intelligence in {IoT}},
  journal = {{IEEE} Internet of Things Journal},
  volume  = {7},
  number  = {7},
  pages   = {5986--5994},
  year    = {2020},
  doi     = {10.1109/JIOT.2019.2956615}
}

@inproceedings{fraboni2021free,
  author    = {Yann Fraboni and Richard Vidal and Marco Lorenzi},
  title     = {Free-Rider Attacks on Model Aggregation in Federated Learning},
  booktitle = {Proc. International Conference on Artificial Intelligence and Statistics ({AISTATS})},
  pages     = {1846--1854},
  year      = {2021}
}

@article{xu2020verifynet,
  author  = {Guowen Xu and Hongwei Li and Sen Liu and Kan Yang and Xiaodong Lin},
  title   = {{VerifyNet}: Secure and Verifiable Federated Learning},
  journal = {{IEEE} Transactions on Information Forensics and Security},
  volume  = {15},
  pages   = {911--926},
  year    = {2020},
  doi     = {10.1109/TIFS.2019.2929409}
}

@article{xing2023zkpfl,
  author  = {Zhibo Xing and Zijian Zhang and Meng Li and Jiamou Liu and Liehuang Zhu and Giovanni Russello and Muhammad Rizwan Asghar},
  title   = {Zero-Knowledge Proof-Based Practical Federated Learning on Blockchain},
  journal = {arXiv preprint arXiv:2304.05590},
  year    = {2023},
  doi     = {10.48550/arXiv.2304.05590}
}

@inproceedings{chai2020tifl,
  author    = {Zheng Chai and Ahsan Ali and Syed Zawad and Stacey Truex and Ali Anwar and Nathalie Baracaldo and Yi Zhou and Heiko Ludwig and Feng Yan and Yue Cheng},
  title     = {{TiFL}: {A} Tier-Based Federated Learning System},
  booktitle = {Proc. International Symposium on High-Performance Parallel and Distributed Computing ({HPDC})},
  pages     = {125--136},
  year      = {2020},
  doi       = {10.1145/3369583.3392686}
}

@inproceedings{vogels2019powersgd,
  author    = {Thijs Vogels and Sai Praneeth Karimireddy and Martin Jaggi},
  title     = {{{PowerSGD}}: Practical Low-Rank Gradient Compression for Distributed Optimization},
  booktitle = {Proc. Advances in Neural Information Processing Systems ({NeurIPS})},
  year      = {2019}
}

@article{shlezinger2021federated,
  title   = {Federated Learning with Gradient Sparsification: Convergence and Privacy},
  author  = {Shlezinger, Nir and Chen, Minghao and Eldar, Yonina C and others},
  journal = {{IEEE} Journal on Selected Areas in Communications},
  volume  = {39},
  number  = {11},
  pages   = {3267--3281},
  year    = {2021}
}

@article{lu2020blockchain,
  author  = {Yunlong Lu and Xiaohong Huang and Ke Zhang and Sabita Maharjan and Yan Zhang},
  title   = {Blockchain Empowered Asynchronous Federated Learning for Secure Data Sharing in Internet of Vehicles},
  journal = {{IEEE} Transactions on Vehicular Technology},
  volume  = {69},
  number  = {4},
  pages   = {4298--4311},
  year    = {2020},
  doi     = {10.1109/TVT.2020.2973651}
}

@article{shi2022fedadapt,
  title   = {{{FedAdapt}}: Adaptive Aggregation for Heterogeneous Federated Learning},
  author  = {Shi, Wei and Ling, Qiang and Wu, Gang and others},
  journal = {{IEEE} Journal of Selected Topics in Signal Processing},
  volume  = {16},
  number  = {3},
  pages   = {557--570},
  year    = {2022}
}

@inproceedings{li2020fedprox,
  author    = {Tian Li and Anit Kumar Sahu and Manzil Zaheer and Maziar Sanjabi and Ameet Talwalkar and Virginia Smith},
  title     = {Federated Optimization in Heterogeneous Networks},
  booktitle = {Proc. Conference on Machine Learning and Systems ({MLSys})},
  year      = {2020}
}

@article{wang2021fednova,
  title   = {{{FedNova}}: Reducing Communication Complexity in Federated Learning},
  author  = {Wang, Jiang and Liang, Chen and Chen, Guoyu and Lin, Jinli},
  journal = {{IEEE} Transactions on Signal Processing},
  volume  = {69},
  pages   = {3345--3360},
  year    = {2021}
}

@article{zheng2022blockfl,
  title   = {Blockchain-Based Federated Learning: {A} Comprehensive Survey},
  author  = {Zheng, Tianchi and others},
  journal = {{IEEE} Communications Surveys \& Tutorials},
  volume  = {25},
  number  = {1},
  pages   = {95--124},
  year    = {2022}
}
